\documentclass{webofc}
\usepackage[varg]{txfonts}   

\usepackage{subcaption}
\begin{document}
\title{Top FCNC searches at HL-LHC with the CMS experiment}
%
%


\author{\firstname{Petr} \lastname{Mandrik}\inst{1}\thanks{\email{petr.mandrik@ihep.ru}} on behalf of the CMS Collaborations
}

\institute{NRC ``Kurchatov Institute'' -- IHEP, Protvino}

\abstract{%
The Large Hadron Collider is the world's largest and highest center-of-mass energy particle accelerator. 
During the Phase I operation it is expected that the LHC operated at a centre-of-mass energy of 13 TeV will 
deliver to the CMS experiment total integrated luminosity of $\sim$300 fb$^{-1}$ till 2023. 
The High Luminosity LHC upgrade is expected to run at a centre-of-mass energy of 14 TeV 
and will allow ATLAS and CMS to collect integrated luminosities of the order of 300 fb$^{-1}$ per year, 
and up to 3000 fb$^{-1}$ during the HL-LHC projected lifetime of ten years. 
The large expected integrated luminosity enables the exploration of the multi-TeV scale by searches for particles
 with high masses as well as by investigation of processes with very low cross sections such as Flavor-Change Neutral Current
 interactions in top quark sector. In this report we present a proposal for the top quark FCNC searches 
at HL-LHC based on Monte-Carlo simulation of the upgraded CMS detector.
}
\maketitle
\section{Introduction} \label{intro}
\begin{figure}
  \centering
  \includegraphics[width=0.9\linewidth,clip]{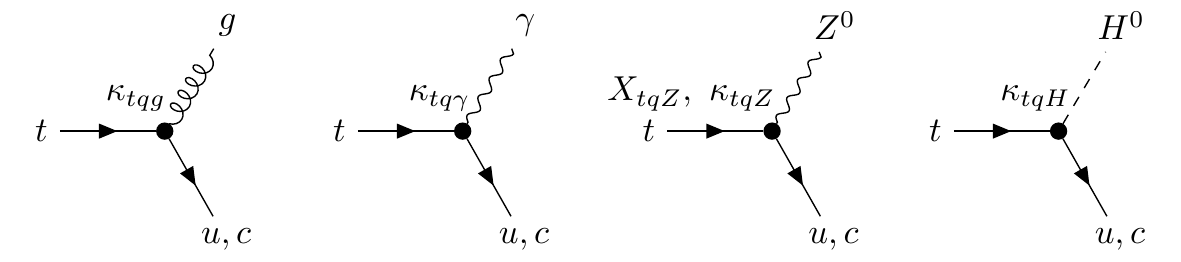}
  \caption{Diagrams for top quark decays mediated by FCNC couplings.}
  \label{fcnc_decays}       
\end{figure}

CMS is a general purpose experiment measuring proton-proton and heavy-ion collisions at the Large Hadron Collider (LHC) at CERN \cite{Chatrchyan:2008aa}.
The detector has been running at proton-proton center-of-mass energies up to 13 TeV, with an LHC luminosity exceeding $10^{34}$ cm$^{-2}$s$^{-1}$.
High Luminosity LHC (HL-LHC) \cite{Apollinari:2116337} is a upcoming major upgrade of the LHC which goal is to integrate ten times more luminosity than the LHC
and allow ATLAS and CMS to collect integrated luminosities of the order of 300 fb$^{-1}$ per year at proton-proton center-of-mass energies of 14 TeV and up to 3000 fb$^{-1}$ during the HL-LHC projected lifetime of ten years.
While the LHC is known as a top factory, the number of produced top quarks will be more than ten times more at HL-LHC due 
to increasing of production cross sections and luminosity. 
It will open new possibility for precise measurement of top-quark properties
as well as allow for probe new physics by studying rare top decays such as those involving flavor changing neutral currents (FCNC, figure~\ref{fcnc_decays})
with upgraded CMS detector \cite{Collaboration:2293646}.


In the Standard Model (SM), FCNC processes are forbidden at tree level and are strongly suppressed in loop corrections by the Glashow-Iliopoulos-Maiani mechanism \cite{PhysRevD.2.1285}.
The predicted SM branching fractions for top quark FCNC decays are expected to be $\mathcal{O}(10^{-12} - 10^{-17})$ \cite{Agashe:2013hma}
and are not detectable at current experimental sensitivity.
However, certain scenarios beyond the SM (BSM), such as
two-Higgs doublet model, warped extra dimensions and minimal supersymmetric models, 
incorporate significantly enhanced FCNC behavior that can be directly probed at the HL-LHC \cite{Agashe:2013hma}.
Observation of such processes would be a clear signal of new physics.

Although flavor-violating couplings of the top may arise from many sources, the effects of BSM physics
in top interactions with a light quark $q = u$, $c$ and a gauge bosons or Higgs may be uniformly described by an effective Lagrangian approach.
The most general effective Lagrangian, containing terms up to dimension five, can be written as \cite{AguilarSaavedra:2004wm}:
\begin{align}
  \begin{aligned}
    -\mathcal{L} & = g_s \kappa_{tqg} \bar{q} 
                     (g_L P_L + g_R P_R)
                     \frac{i \sigma_{\mu\nu} q^\nu}{\Lambda} T^a t G^{a\mu}
                   + e \kappa_{tq\gamma} \bar{q} 
                     (\gamma_L P_L + \gamma_R P_R)
                     \frac{i \sigma_{\mu\nu} q^\nu}{\Lambda} t A^{\mu} + \\
                 & + \frac{g}{2 c_W} X_{tqZ} \bar{q} 
                     (x_L P_L + x_R P_R)
                     t Z^{\mu}
                   + \frac{g}{2 c_W} \kappa_{tqZ} \bar{q} 
                     (z_L P_L + z_R P_R)
                     \frac{i \sigma_{\mu\nu} q^\nu}{\Lambda} t Z^{\mu} + \\
                 & + \frac{g}{2\sqrt{2}} \kappa_{tqH} \bar{q}
                     (h_L P_L + h_R P_R)
                     t H^{\mu} + h.c.,
  \end{aligned}
\end{align}
where $P_L$ and $P_R$ are chirality projectors in spin space, $\kappa_{tqX}$ and $X_{tqZ}$ are effective couplings
for the corresponding vertices, $\Lambda$ is the scale of new physics. 
To avoid ambiguities due to different normalizations of the couplings in the Lagrangian,
the branching ratios of the corresponding FCNC processes are used for a comparison and presentation of experimental results.

\section{Overview of results}
In this section, the study of sensitivity of the upgraded CMS detector to top quark FCNC is presented 
for integrated luminosities of 300 and 3000 fb$^{-1}$ based on Monte-Carlo (MC) simulation. 

\subsection{$t \rightarrow q\gamma$ branching ratios}
The preliminary estimation of $t \rightarrow q \gamma$ branching ratios ($q=u$ or $c$ quark) 
is based on searches of single top quark production via $q \rightarrow t\gamma$ \cite{Collaboration:2293646}.
The final-state signature of signal contain an isolated high energetic photon and products of SM decay of the top.

The signals and $t\bar{t}$ ($+\gamma$), single top and $W+$jets ($+\gamma$) backgrounds are generated
using the full MC simulation of the CMS Phase-2 detector, while the associated production of a single top quark with one or two additional photons
is estimated from fast Delphes simulation \cite{deFavereau2014}.
Background and signal are produced with realistic Phase-2 conditions 
using the pileup scenario corresponding to a mean of 200 additional interactions per single bunch crossing.


Events are selected requiring the presence of exactly one isolated lepton (muon or electron) with $p_T > 25$ GeV and $|\eta| < 2.8$,
at least one isolated photon with $p_T > 50$ GeV and $|\eta| < 2.8$ and
two or three jets with $p_T > 30$ GeV and $|\eta| < 2.8$ reconstructed using the anti-$k_T$ algorithm with a distance parameter $R = 0.4$.
The event is required to have exactly one jet that passes $b$ tagging criteria
with efficiency of about 70\% and a misidentification probability of 18\% for $c$ jets, and 1.5\% for other jets.
Electrons and photons in the overlap region between the barrel and endcaps electromagnetic calorimeters ($1.4 < |\eta| < 1.6$) are removed.
The event is rejected if it has additional muon or electron candidates with $p_T > 10$ GeV and $|\eta| < 2.8$
and if the reconstructed mass of top quark is not in range from 130 to 220 GeV.
In additional all objects are required to be well separated with $\Delta R(e/\mu,\gamma) > 0.7$ and $\Delta R(b,\gamma) > 0.7$.

\begin{figure}[]
  \centering
  \begin{subfigure}[t]{0.49\textwidth}
    \centering
    \includegraphics[width=\linewidth,clip]{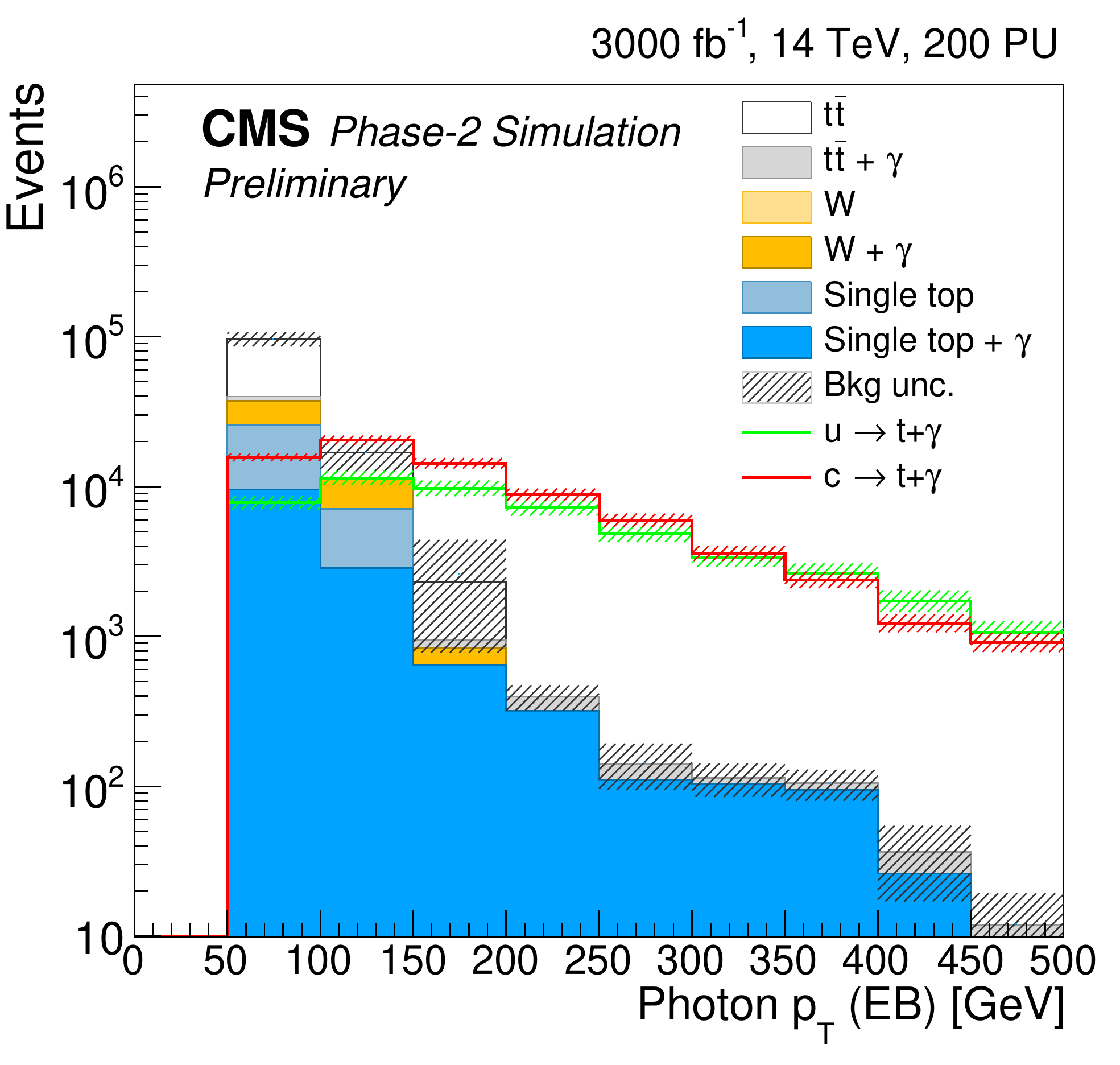}
  \end{subfigure}
  \begin{subfigure}[t]{0.49\textwidth}
    \centering
    \includegraphics[width=\linewidth,clip]{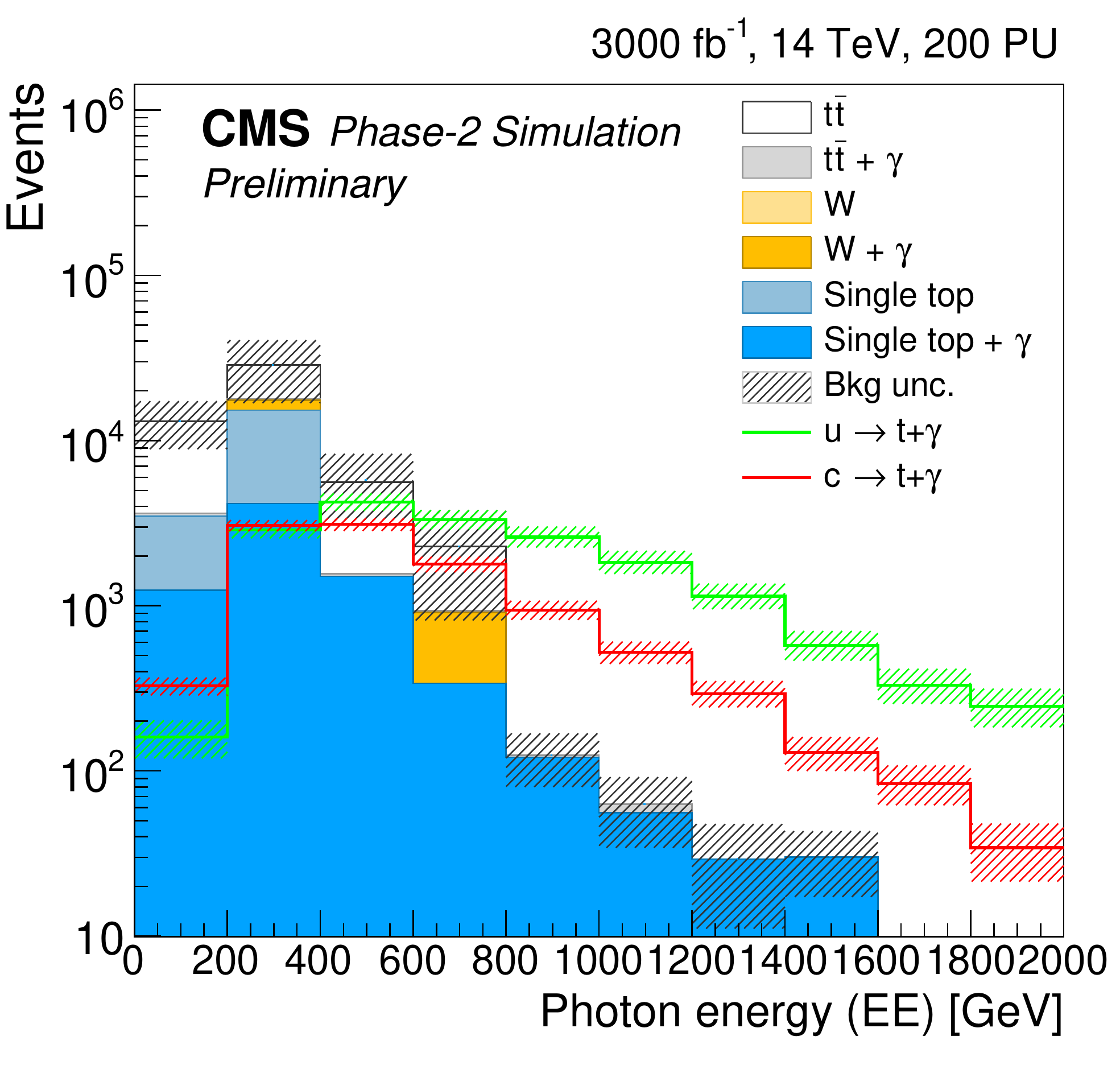}
  \end{subfigure}
  \caption{Photon candidates distributions of transverse momentum for the central region ($|\eta| < 1.4$) (left) and 
           energy in the forward region ($1.6 < |\eta| < 2.8$) (right).
           For illustrative purposes, the signal samples are normalized to production cross section of 1 pb \cite{Collaboration:2293646}.
}
  \label{fig_tqgamma}       
\end{figure}

The region with high $p_T$ (energy) of the photon candidate is populated by signal events,
while the low $p_T$ (energy) region is background dominated (figure~\ref{fig_tqgamma}).
This feature is used in order to obtain the discriminating distribution for the statistical analysis.
Following sources of uncertainties are treated as nuisance parameters in the fit to the background-only model:
the uncertainty in the $b$ tagging efficiency (1\% for $b$ jets, 2\% for $c$ jets and 5\% for light jets), 
jet energy scale (1\%), 
lepton efficiencies (1\%), 
total integrated luminosity (1.5\%),
$t\bar{t}$ cross section (6\%),
single top + photon cross section (30\%),
and the uncertainties in cross sections of the remaining processes according to variation of the renormalization and factorization scales.
The expected 95\% confidence level upper limits on the branching fractions
for an integrated luminosity of 3000 fb$^{-1}$ are found to be 
$\mathcal{B}(t \rightarrow u\gamma) < 0.9 \times 10^{-3} \%$ and 
$\mathcal{B}(t \rightarrow c\gamma) < 7.4 \times 10^{-3} \%$.
A comparison of the results with exclusion  limits  obtained  from previous searches is presented in table~\ref{tqgamma_table}.

\begin{table}[h]
  \centering
  \caption{ $t \rightarrow q\gamma$ branching ratio upper limits at 95\% C.L. from single top quark production analyzes.}
  \label{tqgamma_table}
  \renewcommand{\arraystretch}{1.5}
  \begin{tabular}{c|c c r}
  \hline
  \hline
  Detector                              & $\mathcal{B}(t \rightarrow u\gamma)$   & $\mathcal{B}(t \rightarrow c\gamma)$  & Ref. \\ \hline
  CMS (19.8 fb$^{-1}$, 8 TeV)          & $0.013 \%$                             & $0.17 \%$                              & \cite{Khachatryan:2015att} \\
  CMS Phase-2 (300 fb$^{-1}$, 14 TeV)   & $0.0021   \%$                          & $0.015 \%$                            & \cite{Collaboration:2293646} \\
  CMS Phase-2 (3000 fb$^{-1}$, 14 TeV)  & $0.0009 \%$                            & $0.0074 \%$                           & \cite{Collaboration:2293646} \\
  \hline
  \hline
  \end{tabular}
\end{table}


\subsection{$t \rightarrow qZ$ branching ratios}
The sensitivity of the upgraded CMS Phase-2 detector to $t \rightarrow q Z$ decay ($q =$ $u$ or $c$)
is studied using top quark pair events with the decay chain 
$tt \rightarrow Zq + Wb \rightarrow \ell \ell q + \nu \ell b$ \cite{CMS-PAS-FTR-13-016}.
The event topology includes the presence of three isolated leptons, one light and one b jets.

In the MC samples of the signal and considered backgrounds ($t\bar{t}$ + jets, $t\bar{t}$ + $Z/W$ + jets,
single top + $Z/W$ + jets,
$Z/W$ + jets,
$WW/WZ/ZZ$ + jets), the fast simulation of the CMS Phase-2 detector and
the inclusion of 140 pileup are performed by the Delphes \cite{deFavereau2014}.

Events are selected requiring the presence of two opposite-signed and isolated leptons
having an invariant mass between 78 GeV and 102 GeV, consistent with a Z-boson decay, an extra isolated lepton,
exactly one $b$ tagged jet and at least one light jets with $p_T > 50$ GeV and $|\eta| < 2.4$
and a significant momentum imbalance in the detector $E_T^{miss} > 30$ GeV.
All leptons (electron or muon) must have $p_T > 30$ GeV and $|\eta| < 2.5$ for electrons and
$|\eta| < 2.4$ for muons.
To suppress cosmic rays contribution the $\mu^+ \mu^-$ pair opening angle is required to be smaller than $\pi - 0.05$.
In additional the event is rejected if the reconstructed mass of top quark from SM decay is not in range from 135.5 to 207.5 GeV.

The predicted signal yield after selection is $578$ events while the total number of background events is about $268$.
A 95\% C.L. upper limit on the branching fraction of $t \rightarrow Zq$ is determined
with the modified frequentist approach using the expected number of background events as observed value
assuming no signal contribution.
Systematic uncertainties from
the jet energy scale (3.4\%), $E_T^{miss}$ resolution (3.2\%),
b-tagging efficiency (4.2\%) and cross section ratio of tqZ to Vtt productions (0.8\%)
are estimated based on the study done with 8 TeV data
and are rescaled to the realistic Phase-2 conditions as $\sqrt{19.5 \text{fb}^{-1} / \mathcal{L} }$ by $0.25$.

The expected 95\% confidence level upper limits on the branching fractions
for an integrated luminosity of 3000 fb$^{-1}$ (300 fb$^{-1}$) are found to be 
$\mathcal{B}(t \rightarrow qZ) < 0.010\%$ ($0.027\%$).
A comparison of the results with exclusion  limits  obtained  from previous searches is presented in table~\ref{tqz_table}.

\begin{figure}[]
  \centering
  \begin{subfigure}[t]{0.49\textwidth}
    \centering
    \includegraphics[width=\linewidth,clip]{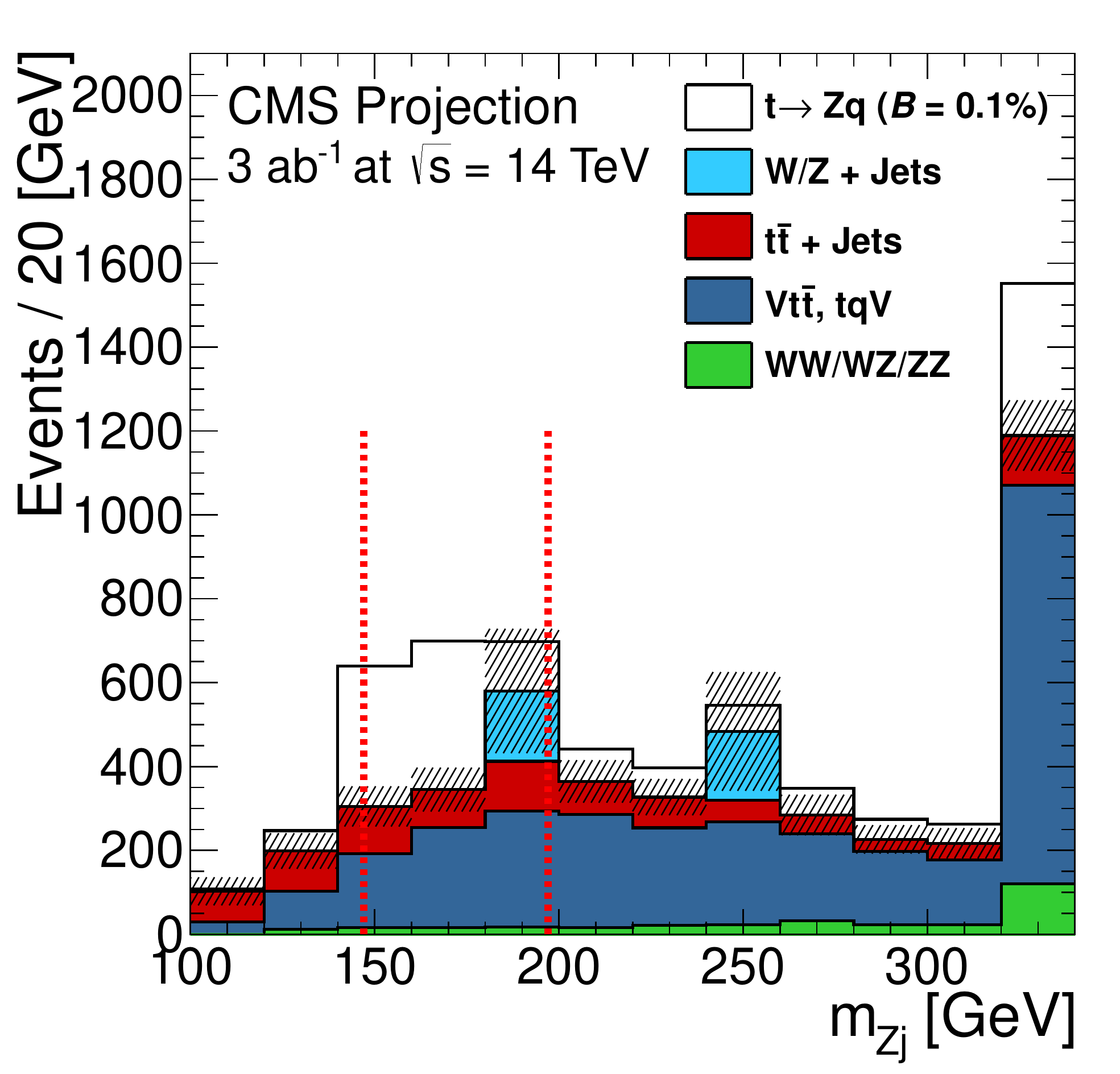}
  \end{subfigure}
  \begin{subfigure}[t]{0.49\textwidth}
    \centering
    \includegraphics[width=\linewidth,clip]{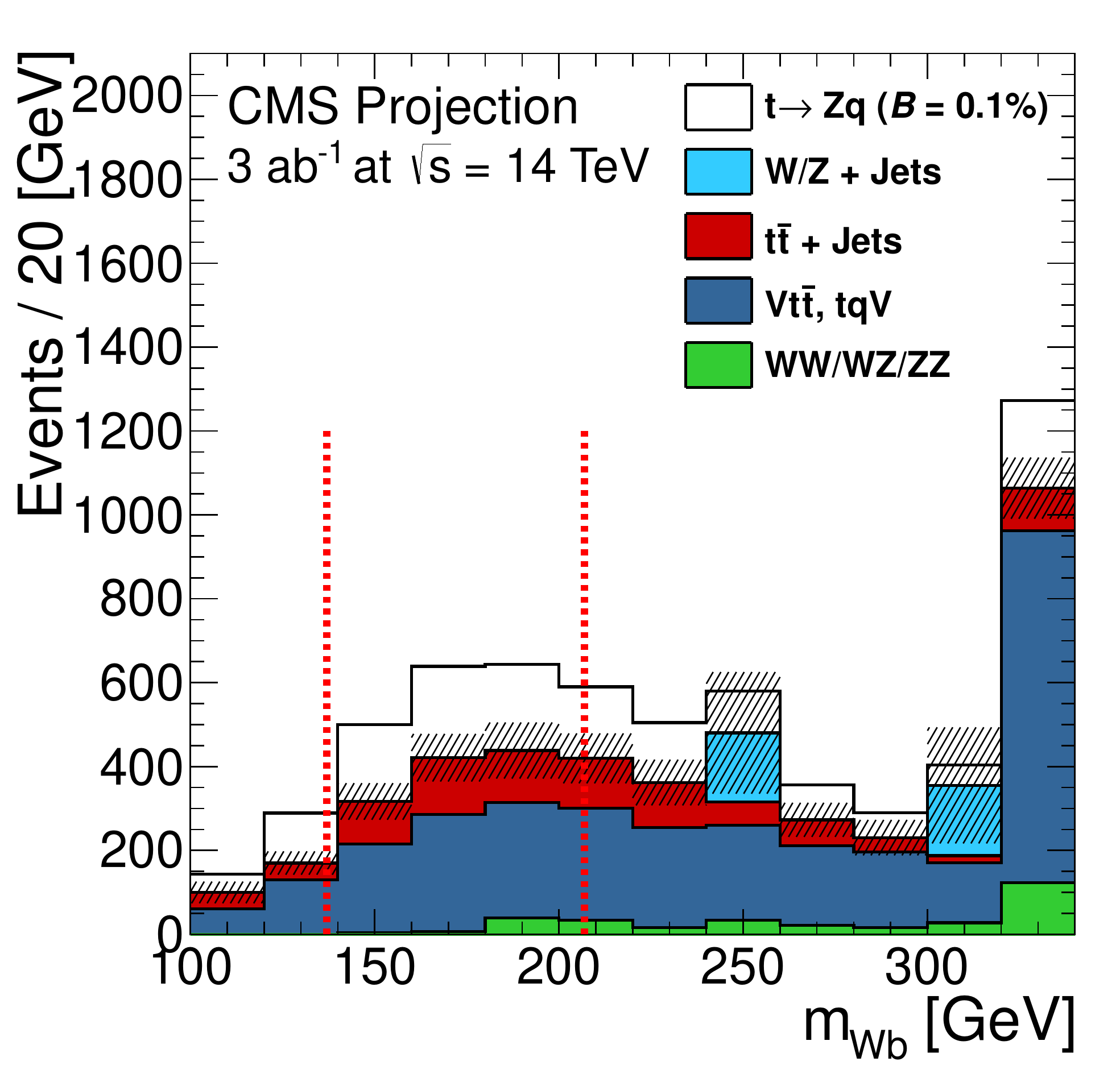}
  \end{subfigure}
  \caption{
  Distributions of reconstructed masses for top quark candidates from decay to $Zq$ via FCNC (left) and from SM decay (right). 
  For illustrative purposes, the signal cross sections are rescaled as $\mathcal{B}(t \rightarrow qZ)$ is equal to 0.1\%.
  The last bins of the two plots contain the overflow events. The dotted vertical lines show the boundaries of the required top mass window \cite{CMS-PAS-FTR-13-016}.
}
  \label{fig_tqZ}       
\end{figure}

\begin{table}[h]
  \centering
  \caption{ $t \rightarrow qZ$ branching ratio upper limits at 95\% C.L. from top pair production analyzes.}
  \label{tqz_table}
  \renewcommand{\arraystretch}{1.5}
  \begin{tabular}{c|c c c r}
  \hline
  \hline
  Detector                              & $\mathcal{B}(t \rightarrow uZ)$ & $\mathcal{B}(t \rightarrow cZ)$ & $\mathcal{B}(t \rightarrow qZ)$  & Ref. \\ \hline
  ATLAS (2.1 fb$^{-1}$, 7 TeV)          & -                                    & -                                    & 0.73\%                                & \cite{Aad:2012ij} \\
  CMS (5.0+19.7 fb$^{-1}$, 7+8 TeV)     & -                                    & -                                    & 0.05\%                                & \cite{Chatrchyan:2013nwa}\\
  ATLAS (36.1 fb$^{-1}$, 13 TeV)        & 0.017 \%                             & 0.024 \%                             & -                                     & \cite{Aaboud:2018nyl}\\
  CMS Phase-2 (300 fb$^{-1}$, 14 TeV)   & -                                    & -                                    & 0.027\%                               & \cite{CMS-PAS-FTR-13-016} \\
  CMS Phase-2 (3000 fb$^{-1}$, 14 TeV)  & -                                    & -                                    & 0.010\%                               & \cite{CMS-PAS-FTR-13-016} \\
  ATLAS (3000 fb$^{-1}$, 14 TeV)        & -                                    & -                                    & 0.0024\% - 0.0058\%                   & \cite{ATL-PHYS-PUB-2016-019} \\
  \hline
  \end{tabular}
\end{table}

\section{Conclusions}
  The HL-LHC will deliver up billions of top quarks,
providing great opportunity to challenge the SM by searching for FCNC processes with upgraded CMS detector.

  Searches for FCNC $tq\gamma$ and $tqZ$ anomalous interactions are projected into new
HL-LHC conditions and shows the possibility of improving existing
constraints on the branchings by about one order of magnitude.
Several studies for different FCNC processes in top quark sector are in an
ongoing state: single top quark production in association with $u$ or $c$ quark via $tqg$ vertex 
and $t \rightarrow qZ$ reanalysis based on a full detector simulation.


Due to huge statistics at the HL-LHC, the systematic uncertainties are expected to be the dominant factor in 
measurement precisions in such analyses.
Therefore, an improvement is required in  accuracy in processes cross section calculations.

\section{Acknowledgments}
I would like to thank L.~Dudko, J.~Kieseler, A.~Savin, M.~Seidel, K.~Skovpen, S.~Slabospitskii and M.~Verzetti for useful discussions.
%
\bibliography{pmandrik_biblio.bib}
%
%

\end{document}